\documentclass[12pt,prd,preprintnumbers]{revtex4}

\usepackage{amsmath,amsfonts}
\usepackage{graphicx}

\newcommand{\id}{{\bf 1}}
\newcommand{\arxiv}[1]{{arXiv:#1}}

\newcommand{\tr}{\mathrm{tr}\,}

\newcommand{\fen}{\mathcal{F}_{\rm eff}}
\newcommand{\arcsinh}{\mathrm{arcsinh}}

\usepackage[normalem]{ulem}  
\usepackage[dvips]{color} 

\renewcommand\sout{\bgroup \color{red} \ULdepth=-.5ex \ULset}

\begin{document}

\preprint{
 RIKEN-MP-48 / 
 YITP-12-47
}

\title{
QCD phase diagram with \\2-flavor lattice fermion formulations
}

\author{Tatsuhiro Misumi}
\email{tmisumi@bnl.gov}
\affiliation{Physics Department, Brookhaven National Laboratory, 
Upton, NY 11973, USA}

\author{Taro Kimura}
\email{tkimura@ribf.riken.jp}
\affiliation{Mathematical Physics Laboratory, RIKEN Nishina Center,
Saitama 351-0198, Japan}

\author{Akira Ohnishi}
\email{ohnishi@yukawa.kyoto-u.ac.jp}
\affiliation{Yukawa Institute for Theoretical Physics, Kyoto University, Kyoto 606-8502, Japan}

\begin{abstract}
We propose a new framework for investigating two-flavor lattice QCD 
with finite temperature and density. We consider the Karsten-Wilczek 
fermion formulation, in which a species-dependent imaginary chemical 
potential term can reduce the number of species to two without losing 
chiral symmetry.
This lattice discretization is useful for study on finite-($T$,$\mu$) QCD 
since its discrete symmetries are appropriate for the case. To show its 
applicability, we study strong-coupling lattice QCD with temperature 
and chemical potential. We derive the effective potential of the scalar 
meson field and obtain a critical line of the chiral phase transition, 
which is qualitatively consistent with the phenomenologically expected 
phase diagram. We also discuss that $O(1/a)$ renormalization of imaginary 
chemical potential can be controlled by adjusting a parameter of a 
dimension-3 counterterm.
\end{abstract}

\maketitle

\newpage


\section{Introduction}
\label{sec:Intro}

Understanding of QCD(Quantum Chromo Dynamics) 
under extreme conditions with temperature and density is
one of keys to elucidating the history of the universe.
In particular, the QCD phase diagram has been attracting 
a great deal of attention (See for example \cite{FH}).
Lattice QCD is the most powerful tool to
investigate such non-perturbative aspects of QCD.
Indeed, the lattice QCD simulation has been applied to the 
finite-temperature QCD with zero density, and has produced 
lots of works to investigate the critical or 
crossover behavior due to confinement and deconfinement 
transition (See for example \cite{Pet}).
However, the Monte-Carlo simulation cannot be easily applied
to QCD with chemical potential because of the notorious sign 
problem (See references in \cite{Forc}).
There have been developed several prescriptions to bypass this problem,
including the imaginary chemical potential method, the Taylor
expansion, the Fugacity expansion and the histogram method 
(See references in \cite{Ejiri}).
Apart from the numerical simulation, the analytical lattice study 
have been also developed.
One of the classical and reliable methods is the strong-coupling
expansion \cite{KaS, KMP, KMNP, DKS, DHK}. 
This method has been applied to the QCD phase diagram 
and has produced successful results
\cite{NFH, Fuk, Nis, KMOO, MNOK, MNO, NMO, MNOK2, NMO2}.
In these works, (unrooted) staggered fermions \cite{KS, Suss, Sha, GS} have 
been used, thus the corresponding continuum theory is 4-flavor 
QCD although the physical two or three-flavor QCD are desirable.

In this paper we propose a new framework of investigating the 
2-flavor finite-($T,\mu$) QCD phase diagram by using the 
Karsten-Wilczek (KW) lattice fermion discretization \cite{KW}. 
This lattice formulation lifts degeneracy of 16 species by introducing 
a species-dependent (imaginary) chemical potential term, instead of introducing a 
species-dependent mass term in Wilson fermion.
The most notable point is that it can reduce
the number of species to 2 with keeping $U(1)$ chiral symmetry. 
It is sometimes called ``minimal-doubling fermions" \cite{KW, CB, CM} 
or ``flavored-chemical-potential(FCP) fermions" \cite{misumi}.
The phase structure in the parameter plane for them has been recently 
studied in \cite{misumi}. In the present work we show that the KW discretization suits 
study on the 2-flavor finite-$(T,\mu)$ QCD phase diagram since it has the same 
discrete symmetries \cite{BBTW, KM, CW} as the finite-density lattice 
QCD system. With progress on the sign problem, this formulation can 
be a powerful tool for the in-medium lattice QCD.
To show the usefulness of the KW fermion, we study strong-coupling 
lattice QCD with temperature and density. We derive the mesonic effective potential 
as a function of ($T$,$\mu$) and elucidate a critical line of the chiral phase 
transition. The result is qualitatively consistent with predictions from 
the phenomenological models. Toward a practical application to lattice 
simulations, we also argue that the $O(1/a)$ renormalization of imaginary chemical 
potential can be controlled by adjusting the relevant parameter $\mu_{3}$ 
of the dimension-3 operator.

In Sec.~\ref{sec:MDF} we investigate the KW fermion and argue
that it is a useful formulation for two-flavor ($T$,$\mu$) lattice QCD.
In Sec.~\ref{sec:SLQ} we study the strong-coupling lattice QCD and 
derive the QCD phase diagram.
Section \ref{sec:SD} is devoted
to a summary and discussion.


\section{Flavored chemical potential}
\label{sec:MDF}

The Karsten-Wilczek (KW) fermion discretization decouples 14 among 
the 16 species in the naive fermion by introducing a species-dependent 
imaginary chemical potential term without losing all the chiral symmetry 
and ultra-locality \cite{KW}. This is a special case of 
``Flavored-chemical-potential (FCP) fermions'' \cite{misumi},
which includes a real-potential type and an imaginary-potential type.
The KW fermion is an imaginary-type FCP fermion and can describe 
two flavors with a proper parameter value, as is called 
``minimal-doubling'' \cite{KW, CB, CM}. 
Since the flavored chemical potential term breaks discrete symmetries 
\cite{BBTW, KM}, we need to fine-tune the three parameters for one dimension-3 
($\bar{\psi}i\gamma_{4}\psi$) and two dimension-4 
($\bar{\psi}\partial_{4}\psi$, $F_{j4}F_{j4}$) counterterms
in order to take a correct Lorentz-symmetric continuum limit for the 
zero-($T$,$\mu$) QCD simulations \cite{CW}.
However, as we will see later, the discrete symmetries of KW fermions 
suit the finite-temperature and -density system, and the severe fine-tuning 
to restore Lorentz symmetry would not be required for this case:
What we need to care about in this case is the dimension-3 operator which
corresponds to an $O(1/a)$ chemical potential term.
This fact inspires us to apply the KW fermion to the in-medium QCD.

\subsection{Symmetries}

\begin{figure}
\includegraphics[height=7cm]{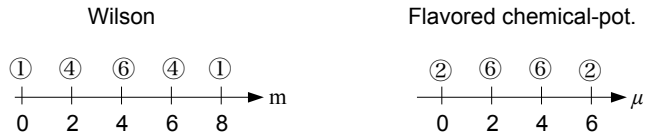} 
\caption{Species-splitting in Wilson and Karsten-Wilczek fermions.
Circled numbers stand for the number of massless flavors on each point.}
\label{MD}
\end{figure}

We study the Karsten-Wilczek fermion and its symmetries in comparison with other 
lattice fermions with chemical potential.
The $4$-d KW action with $U(1)$ chiral symmetry 
and ultra-locality is obtained by introducing a Wilson-like term 
proportional to $i\gamma_4$ into the naive fermion action as 
\begin{eqnarray}
 S_{\mathrm{KW}} & = &
  \sum_{x} 
  \Bigg[
   \frac{1}{2} \sum_{\mu=1}^4 \bar\psi_x \gamma_\mu
   \left(
    U_{x,x+\mu} \psi_{x+\mu} - U_{x,x-\mu}\psi_{x-\mu}
   \right)
   \nonumber \\
 & & + r\frac{i}{2}\sum_{j=1}^3 \bar \psi_x \gamma_4
  \left( 2\psi_{x} - 
    U_{x,x+j} \psi_{x+j} - U_{x,x-j} \psi_{x-j}
   \right) 
   \nonumber\\
  &&
  + i\mu_{3}\bar{\psi}_{x}\gamma_{4}\psi_{x}
  \Bigg],
\label{Smd}  
\end{eqnarray}
where the second line with a parameter $r$ is a 
flavored-chemical-potential term, 
which works to lift the degeneracy of species. 
The third line with a relevant parameter $\mu_{3}$ is 
a dimension-3 counterterm, which 
corresponds to an $O(1/a)$ chemical potential term 
\footnote{
In S.~Kamata, H.~Tanaka, [\arxiv{1111.4536}] (2011)
the authors discussed difficulty of defining hermiticity in the 
minimal-doubling fermions. However it is always the case with 
lattice fermions with (imaginary) chemical potential, which lose the PT invariance. 
Their argument does not mean any problem peculiar to the formulations.}.
If one drops $i$ in front of the second and third lines,
this becomes a real-type Karsten-Wilczek fermion without $\gamma_{5}$
hermiticity while we in this paper focus on the imaginary-type 
KW fermion basically.
For the free theory, the associated Dirac operator in momentum 
space is given by
\begin{equation}
 aD_{\mathrm{KW}}(p) =
  i \sum_{\mu=1}^4 \gamma_\mu \sin ap_\mu
  + i\gamma_4
(\mu_{3}+3r-  r  \sum_{j=1}^{3}\cos ap_j).
 \label{Smdp} 
\end{equation}
For $r=1$ and $\mu_{3}=0$, it has only two zeros at 
$p=(0, 0, 0, 0), (0,0,0,\pi/a)$.
The rest 14 species have $O(1/a)$ imaginary chemical potential 
due to the flavored-chemical-potential term. 
More precisely, among the original 16 species, two have zero imaginary chemical
potential, six have $2/a$, six have $4/a$ and two have $6/a$.
In the naive continuum limit, the 14 species are decoupled 
with infinite imaginary chemical potential and there remains only two flavors as 
shown in Fig.~\ref{MD}
\footnote{These two species are not equivalent since the gamma matrices are 
differently defined between them as $\gamma_\mu'=\Gamma^{-1} \gamma_\mu \Gamma$. 
In the above case it is given by $\Gamma = i \gamma_4 \gamma_5$. 
This means the chiral symmetry possessed by this action is
identified as a flavored one given by $\gamma_5\otimes\tau_3$.}.
The flavored chemical potential term breaks discrete symmetries.
The residual symmetries are cubic symmetry, corresponding to 
permutation of spatial three axes, CT and P \cite{BBTW}.
We list the symmetries of importance as following;
\vspace{0.5cm}

(1) $U(1)$ chiral symmetry ($\gamma_{5}\otimes\tau_{3}$ \cite{T, CKM1, Saidi, Saidi2, CSR})

(2) P

(3) CT

(4) Cubic symmetry. 

\vspace{0.5cm}
Since the cubic symmetry is likely to be enhanced to the 3d rotation
symmetry in the continuum, we expect that these symmetries 
become those of the finite-density QCD in the continuum limit.
To convince ourselves, let us look into symmetries of the naive lattice fermion
with chemical potential, which is introduced as a 4th-direction abelian gauge field.
The massless naive action with chemical potential is given by
\begin{align}
S_{\rm n}(\mu) ={1\over{2}}\sum_{x} \Bigg[
\sum_{j=1}^3 \bar\psi_x \gamma_j
&\left(U_{x,x+j} \psi_{x+j} -U_{x,x-j}\psi_{x-j}
\right) \nonumber\\    
&+\bar\psi_x \gamma_4
\left(e^{\mu}U_{x,x+4} \psi_{x+4} 
-e^{-\mu}U_{x,x-4}\psi_{x-4}\right).
\Bigg]
\label{Smu}
\end{align}
The chemical potential breaks the hypercubic symmetry
into the spatial cubic symmetry. It also breaks C, P and T symmetries into
CT and P. These discrete symmetries are the same as those of 
Karsten-Wilczek fermion as shown above. From the viewpoint of the 
universality class, these two theories belong to the same class. 
We therefore consider that the KW fermion suits in-medium lattice QCD 
much better than zero-($T$,$\mu$) lattice QCD.

However, we have to care about the way of introducing chemical potential.
In KW fermions the flavored imaginary chemical potential is introduced naively
as $\bar{\psi}i\gamma_{4}\psi$ while the chemical potential is usually introduced 
as a 4th-direction abelian gauge field on the lattice as shown in (\ref{Smu}). 
As is well-known \cite{HK}, the naive introduction of chemical potential
violates the abelian gauge invariance and requires a counterterm to
make the energy density and other thermodynamical quantities finite. 
We can discuss the necessity of a counterterm also from the viewpoint of
the additive renormalization: We remind ourselves that the Wilson fermion breaks 
chiral symmetry by $O(1/a)$ flavored-mass terms, which leads to the additive 
mass renormalization and the necessity of the mass parameter tuning 
in the interacting theory. Now the KW fermion breaks discrete 
symmetries into those of finite-density systems by $O(1/a)$ flavored-chemical-potential terms.
These facts indicate that we will here encounter large chemical 
potential renormalization instead of the additive mass renormalization in Wilson fermion.
This is why we need to introduce the dimension-3 counterterm as 
$\mu_{3}\bar{\psi}i\gamma_{4}\psi$ in the action (\ref{Smd}) even for the application 
to finite-density QCD.
The zero-chemical-potential two flavors for a free case in (\ref{Smdp})
suffer $O(1/a)$ imaginary chemical potential renormalization in the interacting theory. 
We thus need to tune $\mu_{3}$ to control it even if we apply it to the finite-density QCD.

\subsection{Additive chemical potential renormalization}
\label{add}

\begin{figure}
\includegraphics[height=7cm]{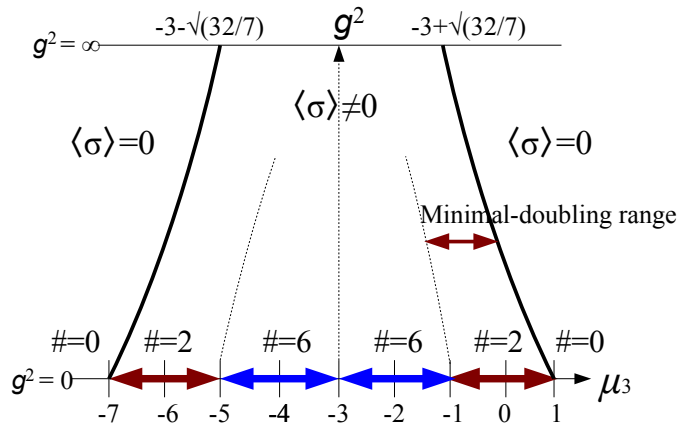} 
\caption{Conjecture on the $\mu_{3}$-$g^{2}$ chiral phase structure 
for the Karsten-Wilczek fermion with r=1.
The width of the minimal-doubling range determines
how hard it is to tune $\mu_{3}$ \cite{misumi}.}
\label{AL}
\end{figure}

As the additive mass renormalization in Wilson fermion is manifested
in the phase diagram in a ($m$--$g^{2}$) plane \cite{AokiP, AokiU1, SS, Creutz3, ACV}, 
the chemical potential renormalization can be manifested in a phase
diagram in a ($\mu_{3}$--$g^{2}$) plane. The chiral phase structure 
in the parameter space of lattice QCD with the KW fermion is studied 
in \cite{misumi} by using strong-coupling lattice QCD and the Gross-Neveu model: 
Fig.~\ref{AL} is the conjectured chiral phase diagram with the number of physical 
flavors in the $(\mu_{3}$--$g^{2})$ plane for $r=1$. There are roughly two phases with and 
without chiral condensate, or equivalently with and without SSB of chiral symmetry.
We name them as ``physical'' and ``unphysical'' phases since the physical QCD has
SSB of chiral symmetry at least.   
As shown in \cite{misumi}, in the strong-coupling and large $N$ limits, 
the boundaries between the two phases are given by
\begin{equation}
\mu_{3}=
\pm{6r^{2}+2\over{\sqrt{6r^{2}+8}}}-3r,
\label{pb}
\end{equation}
which gives the physical range $-\sqrt{32/7}-3<\mu_{3}<+\sqrt{32/7}-3$ for $r=1$. 
We thus have the two chiral boundaries in this limit as shown in Fig.~\ref{AL}.

In the weak-coupling limit ($g^{2}=0$) we analytically know
the number of flavors. In (\ref{Smdp}) with $r=1$, the number of
flavors changes with $\mu_{3}$ being varied \cite{misumi}: 
There are four sectors with two, six, six and two flavors as shown in Fig.~\ref{AL}.
(On boundaries between the sectors $\mu_{3}=1,-1,-3,-5,-7$, 
fermions have unusual dispersions as $D(p)\sim {\bf p}+p_{4}^{2}$, which cannot be
fixed even by tuning parameters. We therefore avoid these points.)
There are no fermion flavors outside these four sectors,
but only unphysical fermions with $O(1/a)$ chemical potential.

As seen from Fig.~\ref{AL}, the boundaries between physical 
$\langle\sigma\rangle\not=0$ and unphysical $\langle\sigma\rangle=0$
phases start from boundaries between the two-flavor and no-flavor sectors in the 
weak-coupling limit. It is reasonable since SSB of chiral symmetry can take place
only in theories with fermions. 
We especially call the two-flavor range ``minimal-doubling range''.
For $r=1$ the minimal-doubling range is given by $-1<\mu_{3}<1$ and $-7<\mu_{3}<-5$.
Toward the strong coupling limit, these ranges are expected to change with 
$g^{2}$ as Fig.~\ref{AL}. 
We note that the minimal-doubling range and the 6-flavor range become
less distinguishable with the gauge coupling being larger.

From the viewpoints of practical application,
the relevant parameter $\mu_{3}$ has to be tuned
to cancel the $O(1/a)$ imaginary chemical potential 
renormalization for the two flavors as discussed in the previous subsection.
In the weak-coupling limit, it is obvious that the two flavors feel no 
chemical potential as long as we set $\mu_{3}$ in the minimal-doubling 
range as $-1<\mu_{3}<1$ or $-7<\mu_{3}<-5$ for $r=1$.
We consider that it carries over in the interacting theory:
For a given gauge coupling, in order to cancel the $O(1/a)$ effective 
chemical potential for the two flavors, we have to set a value of 
$\mu_{3}$ in the minimal-doubling range.
As conjectured in Fig.~\ref{AL}, the minimal-doubling range 
for the middle gauge coupling is likely to have some width, which 
means that we do not need to fine-tune $\mu_{3}$, just set it in the range.
After we tune $\mu_{3}$ as such, there will remain only $O(1)$ 
renormalization of imaginary chemical potential which has a physical scale.

From these arguments, it becomes clear that KW fermion 
can be applied to study on the finite-($T$,$\mu$) QCD as long as
$\mu_{3}$ is chosen to be a proper value.
We here write the KW action with a usual chemical potential 
parameter $\mu$ as
\begin{align}
 S_{\mathrm{KW}}(\mu) & = 
  \sum_{x} 
  \Bigg[
   \frac{1}{2} \sum_{j=1}^3 \bar\psi_x \gamma_j
   \left(
    U_{x,x+j} \psi_{x+j} - U_{x,x-j}\psi_{x-j}
   \right)
   \nonumber\\
   &+{1\over{2}}\bar\psi_x \gamma_4
   \left(
    e^{\mu}U_{x,x+4} \psi_{x+4} 
    - e^{-\mu}U_{x,x-4}\psi_{x-4}\right)  
   \nonumber \\
 &+ r\frac{i}{2}\sum_{j=1}^3 \bar \psi_x \gamma_4
  \left( 2\psi_{x} - 
    U_{x,x+j} \psi_{x+j} - U_{x,x-j} \psi_{x-j}
   \right) + i\mu_{3}\bar{\psi}_{x}\gamma_{4}\psi_{x}
  \Bigg],
\label{Smdmu}  
\end{align}
where $\mu$ is a chemical potential parameter, which can be real $\mu_{\rm Re}$,
imaginary $i\mu_{\rm Im}$ or complex $\mu_{\rm Re}+i\mu_{\rm Im}$.
As we discussed, we consider that we can keep the additive $O(1/a)$ 
renormalization of imaginary chemical potential under control 
by setting $\mu_{3}$ within the range.
We also need to adjust $\mu_{\rm Im}$ to control the $O(1)$ imaginary chemical potential. 
We expect we can be informed of the size of the effective imaginary chemical potential 
by $\pi_4$ condensate in small $T$ and $\mu_{\rm Re}$ region
as we will discuss later.

In the end of this section, we give a comment on practical applications 
of KW fermions to numerical simulations.
Since the present lattice simulation can be applied only to
systems with small $\mu_{\rm Re}$ or with $\mu_{\rm Im}$,
the most feasible application of the KW fermion for now could be the 
imaginary-chemical-potential lattice QCD.
In such a case we can in principle describe two flavors with 
arbitrary $\mu_{\rm Im}$ by keeping $\mu_{3}$ within the minimal-doubling 
range and adjusting $\mu_{\rm Im}$ properly.


\section{Strong-coupling lattice QCD}
\label{sec:SLQ}

To show that the Karsten-Wilczek formulation works in the
study of finite-temperature and finite-density lattice QCD, 
we study the QCD phase diagram in the framework of 
strong-coupling lattice QCD with this formulation.
The strong-coupling lattice QCD study with 
minimal-doubling fermions
has been first performed in \cite{Rev},
where spontaneous chiral symmetry breaking due to chiral condensate 
is observed. We extend this to finite-temperature and finite-density cases
by following a method in \cite{NFH, Fuk, Nis},
and elucidate the dependence of chiral condensate on 
temperature and chemical potential. 
We will find a phase structure consistent with the phenomenological models.

Before starting analysis, we give some comments on our analysis:
Firstly, in the strong-coupling limit, the notion of ``species'' gets ambiguous as shown
in Fig.~\ref{AL}: We cannot distinguish 2-flavor and 6-flavor ranges in the $(\mu_{3}, g^{2})$
plane, but we can just distinguish physical ($\langle \sigma \rangle \not= 0$) and 
unphysical ($\langle \sigma \rangle = 0$) regions in the strong-coupling.  
In this section we will just choose a value of $\mu_{3}$ within the physical range,
and obtain the QCD phase diagram. 
Our purpose here is just to show that the KW formulation works to study 
finite-$(T,\mu)$ QCD, thus the analysis with this modest condition is sufficient.
Secondly, as we have discussed, we consider the feasible numerical application of 
the KW fermion would be the imaginary-$\mu$ lattice QCD for now.
However, we in this section introduce real-$\mu$ and study
the finite-($T$, $\mu_{\rm Re}$) phase structure since
the strong-coupling lattice QCD is free from the sign problem and 
the results can be compared to the phenomenological predictions.


\subsection{Effective potential}
 
We now derive the effective potential of the scalar meson field $\sigma$ 
from (\ref{Smdmu}) in the strong-coupling limit ($g^{2}\to\infty$).
We here consider general color number $N_{c}$ for the $SU(N_{c})$ 
gauge group and general space-time dimensions as $d=D+1$. We
first perform the 1-link integral for the gauge field in the 
$D$-dimensional spatial part, and introduce auxiliary fields 
to eliminate the 4-point interactions as
\begin{align}
 \int \mathcal{D} U_1 \cdots \mathcal{D} U_D \, &\exp
  \left[
   - \sum_{x} \sum_{j=1}^D
   \left(
      \bar\psi_x P_{j}^+ U_j(x) \psi_{x+\hat{j}}
    - \bar\psi_{x+\hat{j}} P_{j}^- U_j^\dag(x) \psi_{x}
   \right)
  \right]
  \nonumber \\
 & =  \exp
  \left[
   N_c \sum_{x} 
   \left(
    \sum_{j=1}^D \tr
    \mathcal{M}(x) ({\it P}_{j}^+)^{\mathrm T} \mathcal{M}(x+\hat{j}) ({\it P}_{j}^-)^{\mathrm T}
   \right)
   + \mathcal{O}(1/\sqrt{D})
  \right]
  \nonumber \\
 & =  \int \mathcal{D} \sigma \mathcal{D} \pi_4 \,
  \exp
  \Bigg[
   - N_c \sum_x
   \Bigg(
    D
    \left(
     (1+r^2) \sigma^2 + (1-r^2) \pi_4^2
    \right)
 \nonumber \\
 & \hspace{7.5em}
    - \frac{D}{2} {\rm tr}
    \left( 
     \sqrt{1+r^2} \sigma - i \sqrt{1-r^2} \pi_4 \gamma_4
    \right) \mathcal{M}(x)
   \Bigg)
  \Bigg],
  \label{integral}
\end{align}
with
\begin{equation}
 P_\mu^{\pm} =
  \left\{
   \begin{array}{lr}
    (\gamma_\mu \pm i r \gamma_4) / 2 & (\mu \not= 4), \\
     \gamma_4 / 2 & (\mu = 4),
   \end{array}
   \right.
\end{equation}
where we introduce the mesonic field as
\begin{equation}
 \mathcal{M}^{\alpha \beta}(x) 
  = \frac{1}{N_c} \delta_{ab} \bar\psi_x^{a,\alpha} \psi_x^{b,\beta} .
\end{equation}
We note that two auxiliary fields $\sigma$ and $\pi_{4}$ are required to get rid of
four-fermi interactions in this case. $\pi_{4}$ condensate is related to density, 
and we will discuss it later.
We also note that we dropped the next-leading order of $O(1/\sqrt{D})$ expansions in
(\ref{integral}), which corresponds to a large $D$ limit.
We now have an intermediate form of the effective action as
\begin{eqnarray}
 S_{\rm eff} & = &
  \sum_{x} \Bigg[
	    \frac{1}{2}
	    \left(
	       \bar\psi_x e^{\mu} U_4(x) \gamma_4 \psi_{x+\hat{4}}
	     - \bar\psi_{x+\hat{4}} e^{-\mu} U_4^\dag(x) \gamma_4 \psi_{x}
	    \right)
	    + \bar\psi_x 
	    \left(
	     m \id + i (\mu_3 + D r) \gamma_4
	    \right) \psi_x
    \nonumber \\
 && 
  + N_c D 
  \left(
   (1+r^2) \sigma^2 + (1-r^2) \pi_4^2
  \right)
  + \frac{N_c}{2} D \, {\rm tr}
  \left(
   \sqrt{1+r^2} \sigma - i \sqrt{1-r^2} \pi_4 \gamma_4
  \right) \mathcal{M}(x) 
  \Bigg] .
  \nonumber \\
\end{eqnarray}
We here consider real chemical potential as $\mu=\mu_{\rm Re}$.
We perform Fourier transformation of the temporal direction by introducing 
Matsubara modes as,
\begin{equation}
 \psi_{\tau,\vec{x}} = \frac{1}{\sqrt{N_\tau}}
  \sum_{n=1}^{N_\tau} e^{ik_n \tau} \tilde \psi_{n,\vec{x}}, \qquad
 \bar\psi_{\tau,\vec{x}} = \frac{1}{\sqrt{N_\tau}}
  \sum_{n=1}^{N_\tau} e^{-ik_n \tau} \tilde{\bar \psi}_{n,\vec{x}},
  \qquad
  k_n = \frac{2\pi}{N_\tau} \left(n - \frac{1}{2}\right) .
\end{equation}
We here take the Polyakov gauge.
The link variable in the temporal direction is given by,
\begin{equation}
 U_4(\vec x) = 
  \left(
   \begin{array}{cccc}
    e^{i\phi_1(\vec x)/N_\tau} & & & \\
    & e^{i\phi_2(\vec x)/N_\tau} & & \\ 
    & & \ddots & \\
    & & & e^{i\phi_{N_c}(\vec x)/N_\tau} \\
   \end{array}
  \right), \qquad
  \sum_{a=1}^{N_c} \phi_a (\vec x) = 0 ,
\end{equation}
with $\phi_{a}$ defined as components of gauge fields.
It enables us to calculate fermionic determinant analytically as, 
\begin{eqnarray}
 \det \mathfrak{D}
 & = &
 \prod_{\vec x} \prod_{a=1}^{N_c} \prod_{n=1}^{N_\tau} \det
  \left[
     \left( m + \frac{D}{2} \sqrt{1+r^2} \sigma \right) \id
   + i \gamma_4 
    \left( \sin \bar{k}_n^{(a)} + \mu_3 + Dr - \frac{D}{2} \sqrt{1-r^2}
     \pi_4 \right)
  \right]
  \nonumber \\
  & \equiv &
  \prod_{\vec x} \prod_{a=1}^{N_c} \prod_{n=1}^{N_\tau} \det
  \left[ B + i \gamma_4 A \sin \tilde{k}_n^{(a)} \right]
  \nonumber \\
 & = &
  \prod_{\vec x} \prod_{a=1}^{N_c} \prod_{n=1}^{N_\tau} 
  \left(
   A^2 \sin^2 \tilde{k}_n^{(a)} + B^2
  \right)^2
  \nonumber \\
 & = &
  \prod_{\vec x} A^{4 N_c N_\tau} \prod_{a=1}^{N_c}
  \left(
   2 \cosh N_\tau E + 2 \cos \left(\phi_a - i N_\tau \mu \right)
  \right)^4 ,
\end{eqnarray}
where we define
\begin{equation}
 A^2 = 1 +
  \left(
   \mu_3 + D r - \frac{D}{2} \sqrt{1-r^2} \pi_4
  \right)^2, \qquad
 B = m + \frac{D}{2} \sqrt{1+r^2} \sigma,
\end{equation}
\begin{equation}
 E = \mathrm{arcsinh} \left(\frac{B}{A}\right)
  = \log
  \left[
   \frac{B}{A} + \sqrt{1+\left(\frac{B}{A}\right)^2}
  \right] , 
\end{equation}
with $\bar k_n^{(a)} = k_n + \phi_a / N_\tau - i \mu$, and
$\tilde{k}_n^{(a)}$ is determined by the relation $A \sin
\tilde{k}_n^{(a)} = \sin \bar{k}_n^{(a)} + \mu_3 + Dr - \frac{D}{2}
\sqrt{1-r^2} \pi_4$.
By integrating the temporal gauge field $\phi_{a}$ we derive
\begin{equation}
 \int \mathcal{D} U_4
   \prod_{\vec x} A^{4 N_c N_\tau} \prod_{a=1}^{N_c}
  \left(
   2 \cosh N_\tau E + 2 \cos \left(\phi_a - i N_\tau \mu \right)
  \right)^4
  = \prod_{\vec x} 
  \left[
   \sum_{n\in\mathbb{Z}} \det\left(Q_{n+i-j}\right)_{1\le i,j\le N_c}
  \right],
\end{equation}
\begin{equation}
 Q_n = \int_{-\pi}^{\pi} \frac{d\phi}{2\pi} 
  \left(
   2 \cosh N_\tau E + 2 \cos \theta 
  \right)^4
  e^{-in\phi}, \qquad 
  \theta = \phi - i N_\tau \mu .
\end{equation}
For $N_{c}=3$ these $Q_{n}$ are explicitly given as
\begin{eqnarray}
 Q_0 = 
  2 (8 \cosh^4 N_\tau E + 24 \cosh^2 N_\tau E + 3), &&
 Q_{\pm 1} =
  8 \cosh N_\tau E (4 \cosh^2 N_\tau E + 3) e^{\pm N_\tau \mu},
 \nonumber \\
 Q_{\pm 2} =
  4 (6 \cosh^2 N_\tau E + 1 ) e^{\pm 2 N_\tau \mu}, &&
 Q_{\pm 3} = 
  8 \cosh N_\tau E \, e^{\pm 3 N_\tau \mu},
 \nonumber \\
 Q_{\pm 4} = e^{\pm 4 N_\tau \mu}, &&
 Q_{|n|\ge 5} = 0 . 
\end{eqnarray}
As a result, the effective potential is given by
\begin{eqnarray}
  \mathcal{F}_{\rm eff}(\sigma,\pi_4;m,T,\mu,\mu_3)
  & = & \frac{N_c D}{4} \left(
			(1+r^2) \sigma^2 + (1-r^2) \pi_4^2
			  \right)
  - N_c \log A
  \nonumber \\
 & &
  - \frac{T}{4} \log \left(
   \sum_{n\in\mathbb{Z}} \det\left(Q_{n+i-j}\right)_{1\le i,j\le N_c}
	   \right) .
	   \label{effV}
\end{eqnarray}
Here we redefine the free energy $4\mathcal{F}_{\rm eff} \to
\mathcal{F}_{\rm eff}$ to be consistent with the phenomenological result
as discussed later.
We here show only the calculation result of the determinant part for $N_c=3$,
\begin{eqnarray}
 &&
  \sum_{n \in \mathbb{Z}} \det\left(Q_{n+i-j}\right)_{1\le i,j\le N_c}
  \nonumber \\
   & = & 
   8 \left(
      1 + 12 \cosh^2 \frac{E}{T} + 8 \cosh^4 \frac{E}{T}
     \right)
   \left(
    15 - 60 \cosh^2 \frac{E}{T} + 160 \cosh^4 \frac{E}{T}
    -32 \cosh^6 \frac{E}{T} + 64 \cosh^8 \frac{E}{T}
   \right)
   \nonumber \\
 &&
  + 64 \cosh \frac{\mu_B}{T} \cosh \frac{E}{T}
  \left(
   -15 + 40 \cosh^2 \frac{E}{T} + 96 \cosh^4 \frac{E}{T}
   + 320 \cosh^8 \frac{E}{T}
  \right)
  \nonumber \\
 &&
  + 80 \cosh \frac{2\mu_B}{T}
  \left(
   1 + 6 \cosh^2 \frac{E}{T} + 24 \cosh^4 \frac{E}{T} 
   + 80 \cosh^6 \frac{E}{T}
  \right)
  \nonumber \\
 &&
  + 80 \cosh \frac{3\mu_B}{T}
  \cosh \frac{E}{T}
  \left(
   - 1 + \cosh^2 \frac{E}{T}
  \right)
  + 2 \cosh \frac{4\mu_B}{T},
\end{eqnarray}
with
\begin{equation}
\mu_B = 3 \mu .
\end{equation}

In the case of zero temperature $T=0$, we can
solve the equilibrium condition analytically.
For $D=3$ ($d=4$) with $m=0$ and $r=1$ the free energy is given by
\begin{align}
 \mathcal{F}_{\rm eff}(\sigma,\pi_4;m, T,\mu, \mu_3) = 
  \frac{9}{2}  \sigma^2 & - \frac{3}{2} \log \left( 1 + (\mu_3+3)^2 \right)
\nonumber\\  
  &- \max
  \left\{
   3 \ \mathrm{arcsinh}
   \left(
    \frac{3 \sqrt{2} \sigma}{2\sqrt{1+(\mu_3+3)^2}}
   \right),  \mu_B
  \right\} .
\end{align}
In this case there are two local minima of the free energy as a function
of $\sigma$, $\fen = -\mu_B - \frac{3}{2} \log \left( 1 + (\mu_3+3)^2
\right)$ at $\sigma = 0$ and $\fen = \frac{9}{2} \sigma^2 -
\frac{3}{2} \log \left( 1 + (\mu_3+3)^2 \right) - 3 \arcsinh (3\,
\sigma/\sqrt{2(1+(\mu_3+3)^2)})$ at $\sigma = \sigma_0$.
This $\sigma_0$ satisfies the following gap equation, 
\begin{equation}
 \frac{\partial \fen}{\partial \sigma}\Bigg|_{\sigma=\sigma_0} = 0
  \quad \longrightarrow \quad
 2 \sigma_0^2
  \left[
   1 + \frac{9}{2} \frac{\sigma_0^2}{1+(\mu_3+3)^2}
  \right]
  = \frac{1}{1+(\mu_3+3)^2} .
\end{equation}
Therefore we have
\begin{equation}
 \sigma_0^2 = \frac{1+(\mu_3+3)^2}{9}
  \left[
   \sqrt{1+\frac{9}{(1+(\mu_3+3)^2)^2}} - 1
  \right] .
\end{equation}
Comparing these two local minima, we can show that the global minimum
changes from $\sigma = \sigma_0$ to $\sigma = 0$ at the critical
chemical potential as
\begin{equation}
 \mu_B^{\rm critical}(T=0) =    
  3 \, \mathrm{arcsinh}
   \left(
    \frac{3 \sqrt{2} \sigma_0}{2\sqrt{1+(\mu_3+3)^2}}
   \right)
   - \frac{9}{2} \sigma_0^2 .
   \label{zero_temp_crit_chem_pot}
\end{equation}
This phase transition is of 1st order because the order parameter
$\sigma$ changes discontinuously at this critical chemical potential.
We can also evaluate the baryon density $\rho_B = - \partial \fen /
\partial \mu_B$ at $T = 0$.
It turns out to be empty $\rho_B =0$ when $\mu_B < \mu_B^{\rm critical}$.
On the other hand, when $\mu_B > \mu_B^{\rm critical}$, it is saturated as
$\rho_B = 1$.

\subsection{Phase diagram}

We depict the QCD phase diagram with respect to 
chiral symmetry from the effective potential (\ref{effV}).
We first concentrate on the case with $r=1$ for simplicity.
In this case the free energy is independent of $\pi_4$ although 
it is an artifact of the strong coupling limit.
Thus we simply neglect $\pi_{4}$ and set $\mu_{3}$ 
in the physical parameter range $(-\sqrt{32/7}<\mu_{3}+3<+\sqrt{32/7})$ in the
strong-coupling limit in Fig.~\ref{AL}. We here take $\mu_{3}=-0.9$.
In a massless case $m=0$, the phase boundary of the 2nd order 
chiral phase transition is given by the condition, such that the 
coefficient of $\sigma^2$ in the free energy becomes zero.
When the order of the phase transition is changed from 2nd to 1st, 
the coefficient of $\sigma^4$ as well as $\sigma^2$ should vanish in the free energy
$\mathcal{F}_{\rm eff}(\sigma)$.

\begin{figure}[t]
 \begin{center}
  \includegraphics[width=20em]{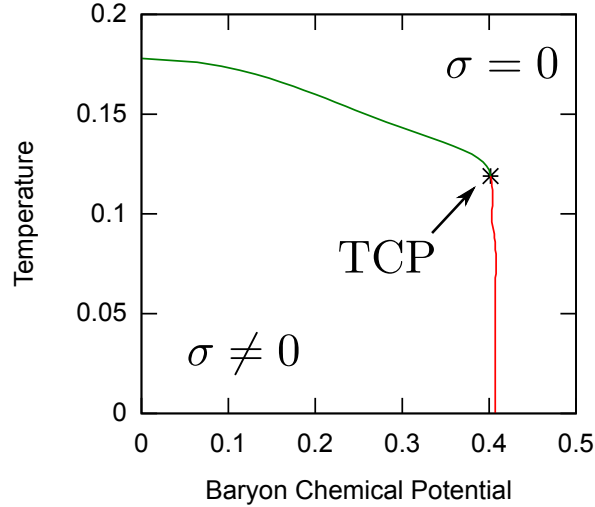}
 \end{center}
 \caption{Phase diagram for the chiral transition with $r=1$, $\mu_3=-0.9$
 and $m=0$. 
 Green and red lines show 2nd and 1st transition lines, respectively.
 The transition order is changed from 2nd to 1st at the tricritical
 point $(\mu_B^{\rm tri},T^{\rm tri})=(0.402,0.119)$.}
 \label{pb_r1}
\end{figure}

\begin{figure}[t]
 \begin{center}
   \includegraphics[width=20em]{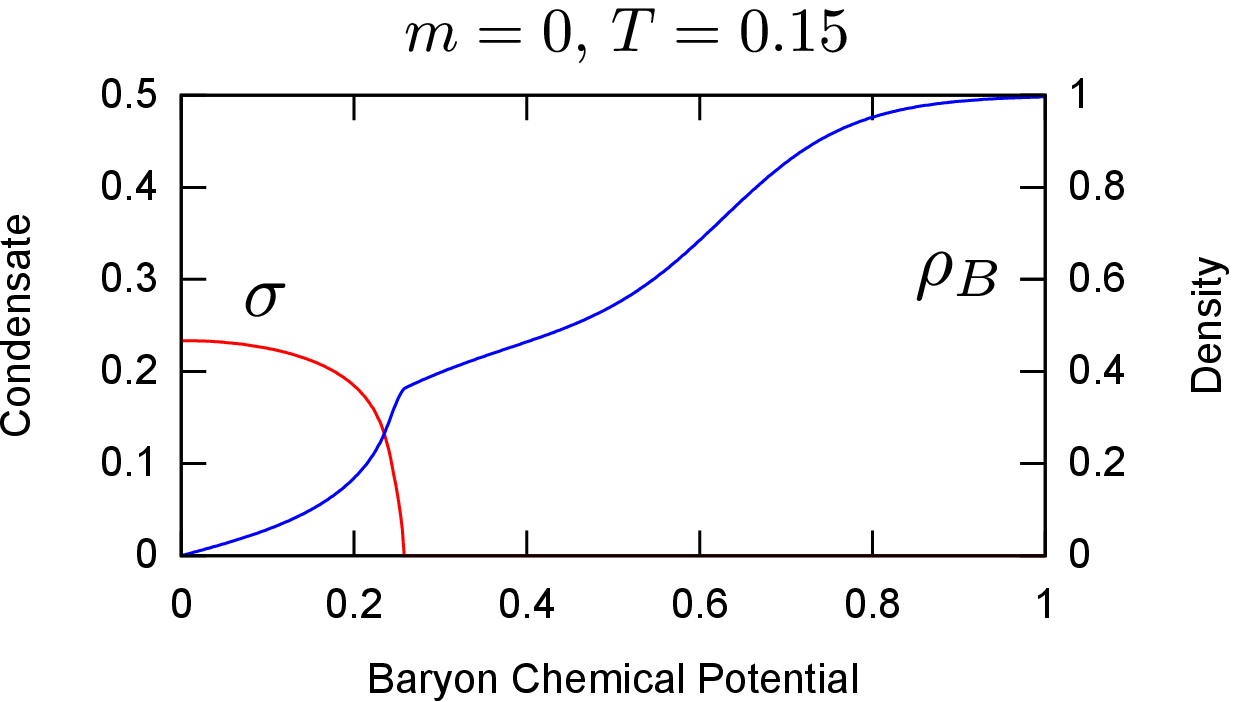} \quad
   \includegraphics[width=20em]{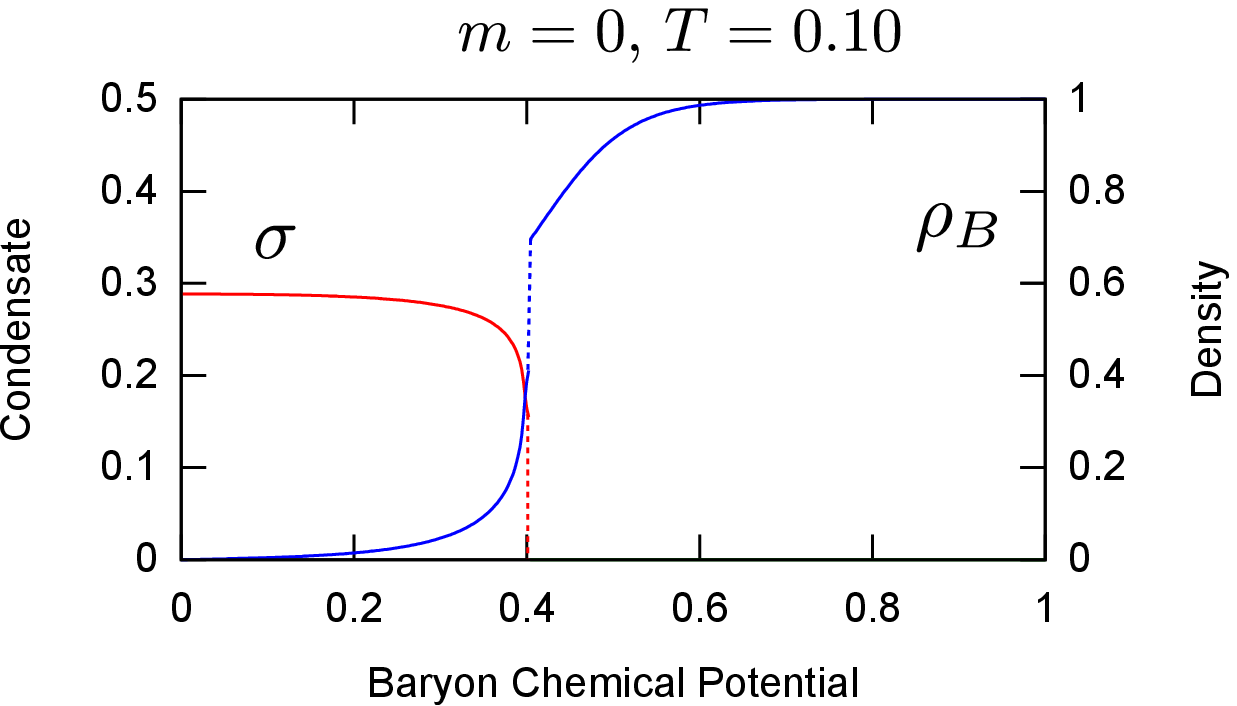} \\ \vspace{1.2em}
   \includegraphics[width=20em]{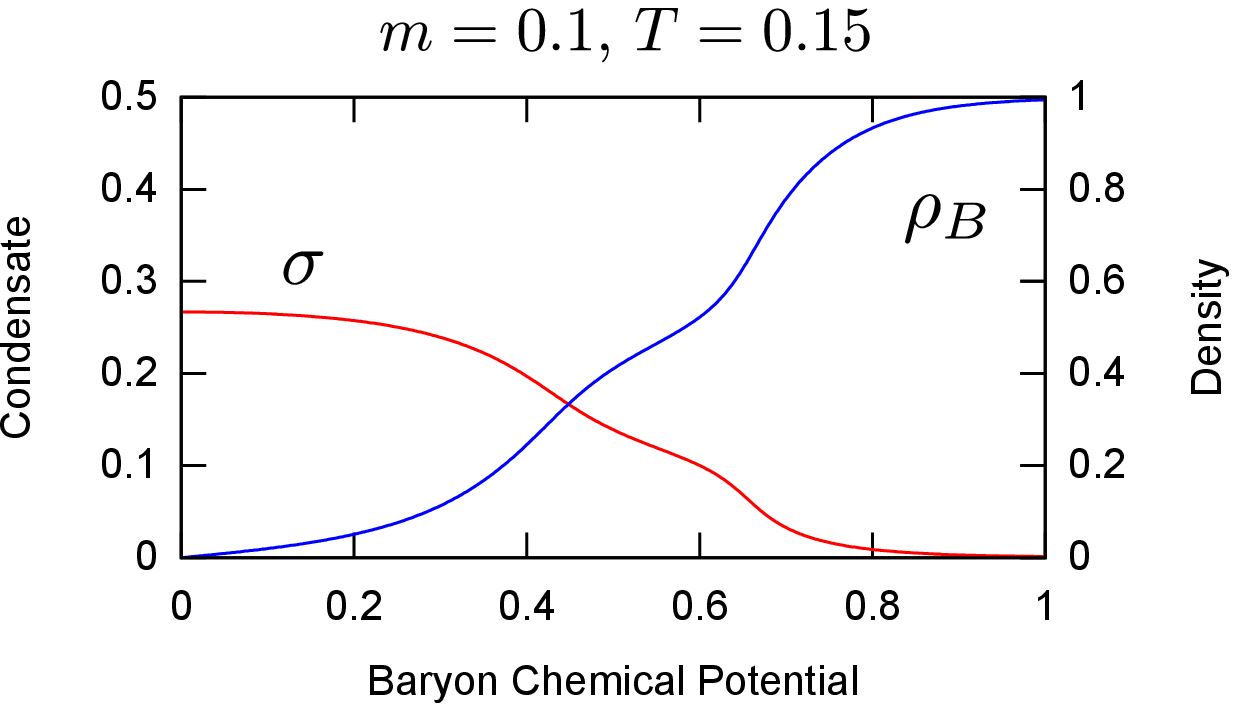} \quad
   \includegraphics[width=20em]{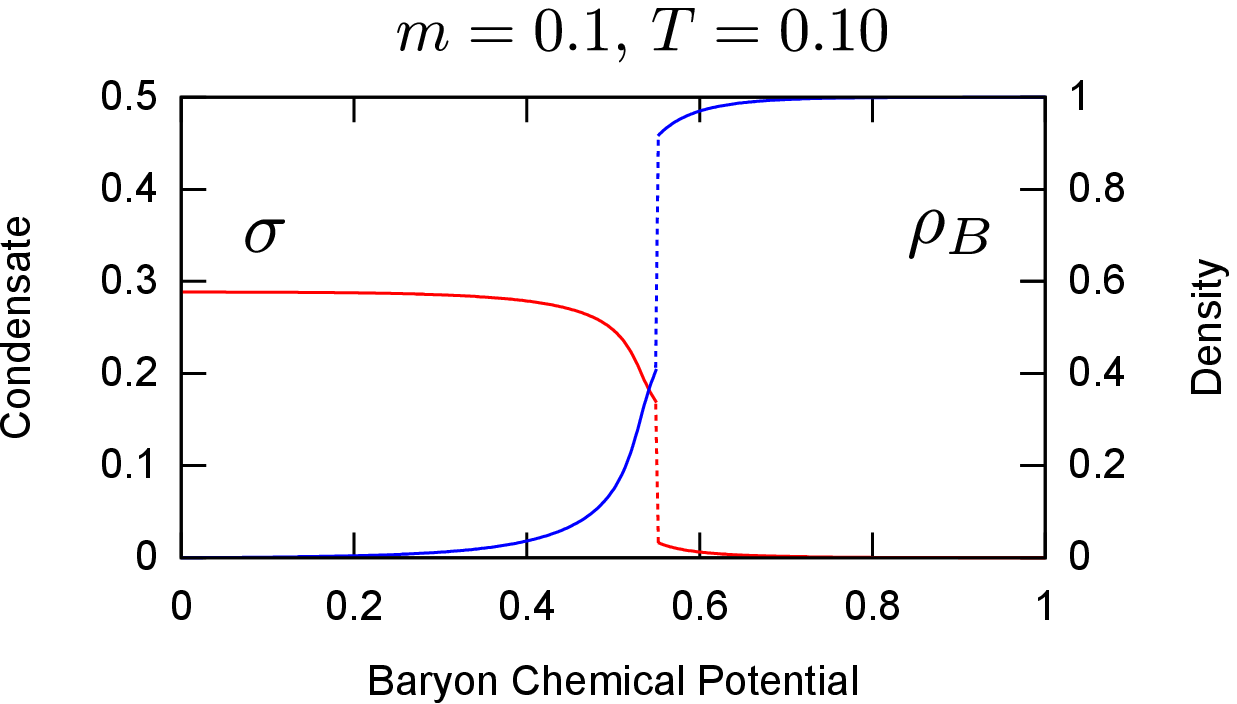}
  \end{center}
 \caption{Chiral condensate $\sigma$ and the baryon density $\rho_B$
 for (left) $T=0.15$ and (right) $T=0.10$.
 Top and bottom panels show the massless $m=0$ and massive $m=0.1$ cases.
 There are 1st and 2nd phase transitions for $\sigma$.
 In the case of $m\not=0$, there appears the crossover behavior instead of the 2nd order transition.
 }
 \label{condensate_r1}
\end{figure}

Figure~\ref{pb_r1} shows the phase boundary of the chiral transition with
$r=1$, $\mu_{3}=-0.9$ and $m=0$.
The order of the phase transition is changed from 2nd to 1st at the
tricritical point $(\mu_B^{\rm tri},T^{\rm tri})=(0.402,0.119)$.
We also depict $\sigma$ condensate and the baryon density 
$\rho_B=-\partial \mathcal{F}_{\rm eff}/\partial \mu_{B}$
as functions of $\mu_{B}$ with several fixed $T$ in Fig.~\ref{condensate_r1}.
We find that there are first ($T<T^\mathrm{tri}$)
and second ($T>T^\mathrm{tri}$) order phase transitions for $\sigma$, 
followed by the phase transition of the density $\rho_B$. 
For $m\not=0$, we can easily show that the crossover transition instead appears 
with the 2nd-order critical point.

These results are qualitatively consistent with those with 
strong-coupling lattice QCD with staggered fermions,
while there are some quantitative differences.
For example, the KW phase diagram is suppressed in $T$ direction 
compared to that in staggered.
We here compare the ratio of the transition baryon chemical potential at $T=0$
to the critical temperature at $\mu_B=0$, $R^{0}=\mu_c(T=0)/T_c(\mu_B=0)$.
In staggered fermion, this ratio is
$R^0_\mathrm{st} \simeq 3 \times 0.56 / (5/3) \sim 1$~\cite{Fuk,Nis},
while $R^0_\mathrm{KW} \simeq 0.406 / 0.178 \sim 2.3$.
In the real world, this ratio is larger,
$R^0 \gtrsim M_N / 170~\mathrm{MeV} \sim 5.5$.
When the finite coupling and Polyakov loop effects are taken into account
for staggered fermion,
$T_c(\mu_B=0)$ decreases, $\mu_c(T=0)$ stays almost constant,
then $R^0$ value increases~\cite{MNOK,MNO,NMO}.
Larger $R^0$ with KW fermion in the strong coupling limit
may suggest smaller finite coupling corrections in the phase boundary.
Another interesting point is the location of the tricritical point.
In KW fermion, the ratio $R^\mathrm{tri}_\mathrm{KW}=0.402/0.119 \simeq 3.4$,
while $R^\mathrm{tri}_\mathrm{st}=1.73/0.866 \simeq 2.0$
for unrooted staggered fermion~\cite{Fuk,Nis}.
It would be too brave to discuss this value,
but $R^\mathrm{tri}_\mathrm{KW}$ is consistent
with the recent Monte-Carlo simulations~\cite{MC},
which implies that the critical point does not exist
in the low baryon chemical potential region, $\mu_B/T \lesssim 3$.
These observations reveal
usefulness of KW fermion for research on QCD phase diagram. 

The $\mu_B$ dependence of $\sigma$ and $\rho_B$ seems to show 
there are two sequential transitions with increasing $\mu_B$.
At $T=0.15 > T^\mathrm{tri}$, $\sigma$ quickly decreases
and $\rho_B$ increases at $\mu_B\simeq 0.25$ $(0.41)$ for $m=0$ $(0.1)$,
and at a larger $\mu_B$ ($\mu_B\simeq 0.64$), 
increasing rate of $\rho_B$ as a function of $\mu_B$ becomes higher again.
At lower temperature, $T=0.10 < T^\mathrm{tri}$,
partial restoration of the chiral symmetry is seen
before the first order phase transition.
Since we have not taken care of the diquark condensate,
these transitions are not related to the color superconductor.
Other types of matter, such as quarkyonic matter~\cite{McLerran},
partial chiral restored matter~\cite{MNOK,MNO}, or nuclear matter, 
may be related to the above sequential change.

\begin{figure}[t]
 \begin{center}
\includegraphics[width=20em]{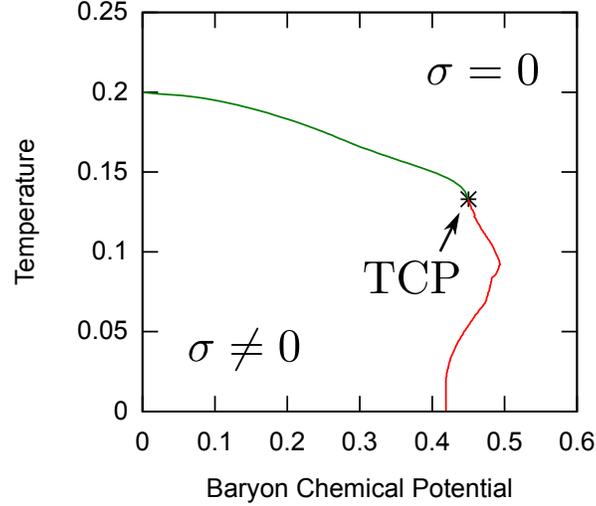}
 \end{center}
 \caption{Phase diagram for the chiral transition with $r=0.75$, $\mu_3=-0.9$
 and $m=0$. 
 The transition order is similarly changed from 2nd to 1st at the tricritical
 point $(\mu_B^{\rm tri},T^{\rm tri})= (0.450,0.133)$.}
 \label{pb_r075}
\end{figure}

\begin{figure}[t]
 \begin{center}
  \includegraphics[width=20em]{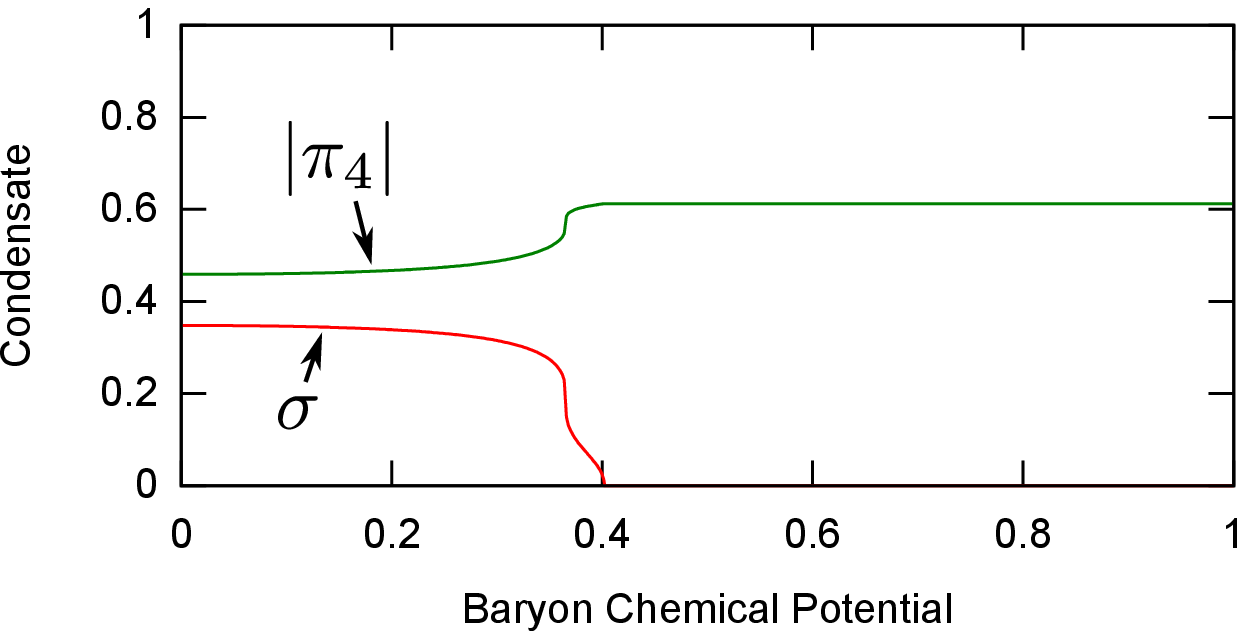} \quad
  \includegraphics[width=20em]{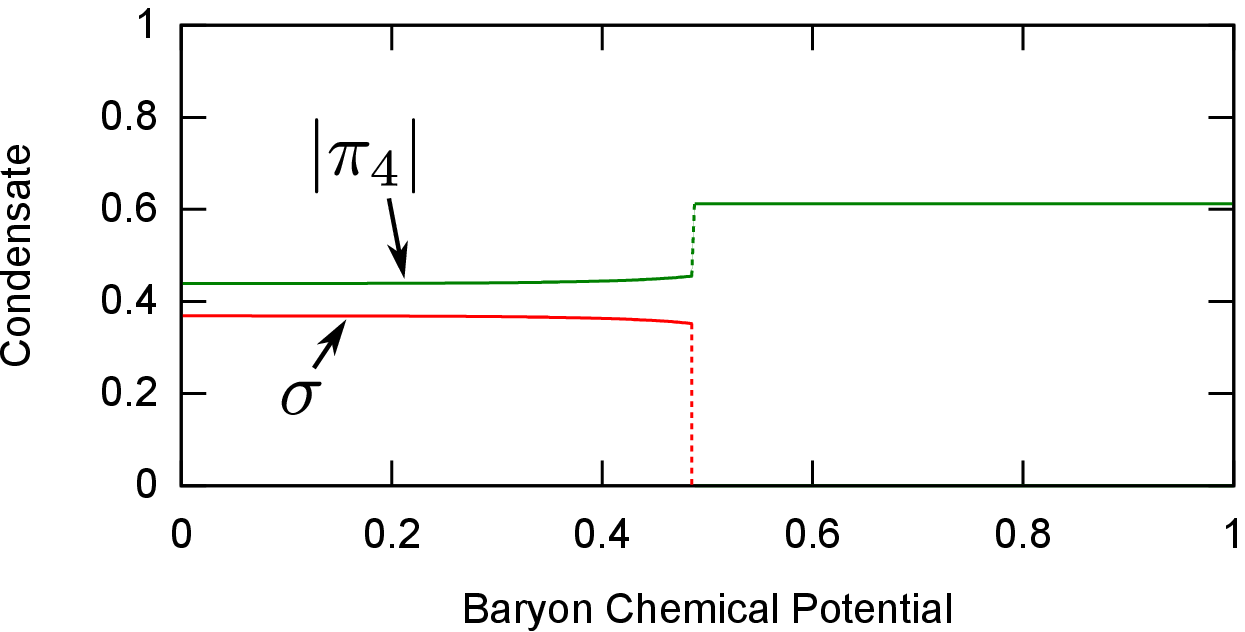}
 \end{center}
 \caption{Condensates $\sigma$ and $|\pi_4|$ for
 (left) $T=0.15$ and (right) $T=0.10$ in the case of $r=0.75$.
 There are 1st and 2nd phase transitions for $\sigma$.}
 \label{condensate_r075}
\end{figure}

We next consider $r\not=1$.
We as an example take $r=0.75$ and $\mu_{3}=-0.9$,
which is again within the physical range.
(For $r=0.75$, the physical parameter range is given by 
$-3.84\lesssim\mu_{3} \lesssim-0.66$ from (\ref{pb}).)
Fig.~\ref{pb_r075} shows the phase diagram for the chiral transition,
which is similar to that of $r=1$ case.
We have a little higher critical temperature, $T_c(\mu_B=0) \simeq 0.20$,
and a little larger transition chemical potential, $\mu_c(T=0) \simeq 0.42$,
compared with $r=1$ case.
The tricritical point, $(\mu_B^\mathrm{tri},T^\mathrm{tri}) \simeq
(0.450,0.133)$, also has a larger $T$ and $\mu_B$.
One difference is the existence of the region, where $d\mu_B/dT>0$ on the first order
transition boundary. This is a lattice artifact, and can be removed when we take account
of finite coupling effects.
The Clausius-Clapeyron relation tells us that
the balance of the pressure in the coexiting two phases, I and II, leads
to the slope of the phase boundary,
$d\mu_B/dT = - (s_\mathrm{II}-s_\mathrm{I})/(\rho_\mathrm{II}-\rho_\mathrm{I})$,
where $s_i$ and $\rho_i$ represent the entropy and baryon density in the phase $i$~\cite{Nis}.
The positive slope of the phase boundary thus implies that the entropy density
in quark matter (higher $\mu_B$ phase) is smaller than that in hadronic matter
(lower $\mu_B$ phase). Since the QCD phase transition is essentially the change of 
the degrees of freedom, we do not expect this to take place in continuum theory.
In the strong copling lattice QCD, however, the baryon density is almost one
in quark matter at low $T$, as discussed at $T=0$ in the previous section.
In the maximum density case ($\rho_B=1$), all the sites are filled by $N_c$ quarks
and the entropy is zero.
Thus the entropy density can be smaller in quark matter, while this is an artifact
of finite spacing lattice.
The same behavior is observed in staggered fermions~\cite{Fuk,Nis},
and it is found that the region with the positive slope boundary narrows or disappears
with finite coupling effects in staggered fermion~\cite{MNO,MNOK,NMO}.

We note that in $r\not=1$ case, we have nonzero $\pi_{4}$ condensate
as shown in Fig.~\ref{condensate_r075}, where we show $\sigma$ and $\pi_{4}$ as
functions of $\mu_{B}$ for several fixed $T$. This $\pi_{4}$ condensate 
also undergoes a phase transition with the chiral transition.
Since $\pi_{4}$ is an auxiliary field for the operator $i\bar{\psi}\gamma_{4}\psi$, 
an absolute value of this vector condensate is deeply related to density.
The point is that this condensate still remains for $\mu_B=0$. 
It is quite natural since the $O(1)$ renormalized imaginary chemical
potential $\tilde{\mu}_{\rm Im}$ due to quantum effects 
can survive even if we set $\mu_{3}$ in the 
minimal-doubling range to eliminate the $O(1/a)$ imaginary chemical potential,
as we discussed in Sec.~\ref{add}.
Thus, the physical imaginary chemical potential is given by the sum of 
the effective one $\tilde{\mu}_{\rm Im}$ and a usual $O(1)$ parameter $\mu_{\rm Im}$
as $\mu_{\rm Im}^{phys}=\tilde{\mu}_{\rm Im}+\mu_{\rm Im}$.
It suggests that one possible way to control $\mu_{\rm Im}^{phys}$ in lattice QCD is 
to check the size of $\pi_{4}$ condensate.

\begin{figure}[t]
 \begin{center}
  \includegraphics[width=25em]{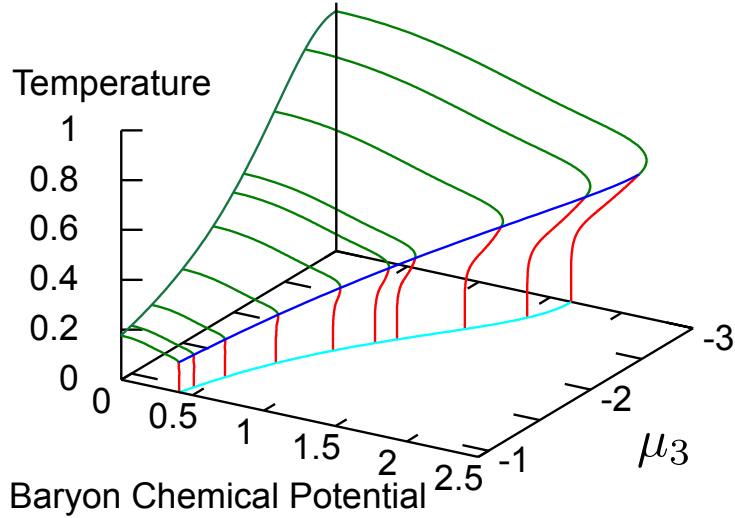}
 \end{center}
 \caption{Three-dimensional chiral phase diagram for $T$, $\mu_B$ and
 $\mu_3$ for $m=0$ where $\mu_{3}$ runs within half of the physical
 range $-3<\mu_3<\sqrt{32/7}-3$. 
 Green, red and purple lines show 2nd, 1st order transitions and
 tricritical point, respectively. The critical chemical potential at $T=0$ is explicitly given by
 (\ref{zero_temp_crit_chem_pot}).}
 \label{pb_3d}
\end{figure}

The existence of $\tilde{\mu}_{\rm Im}$ also affects the $\mu_{3}$ dependence of critical lines:
In Fig.~\ref{pb_3d} we change $\mu_{3}$ and depict a three-dimensional chiral phase 
diagram for $T$, $\mu_B$ and $\mu_{3}$ for $r=1$. It shows that the critical line changes with 
$\mu_{3}$ being varied. As far as we keep $\mu_{3}$ in the physical range, 
the $O(1/a)$ effective chemical potential for the physical flavors is cancelled. 
As discussed above, however, we still have $O(1)$ contribution as $\tilde{\mu}_{\rm Im}$. 
We can interpret that the dependence of the critical line on $\mu_{3}$ comes from
the dependence of $\tilde{\mu}_{\rm Im}$ on $\mu_{3}$ as $\tilde{\mu}_{\rm Im}(\mu_{3})$.

In this section we have obtained the finite-($T$,$\mu$) QCD phase diagram in the 
strong-coupling limit. To be precise, since the theory effectively contains the 
renormalized imaginary chemical potential as the KW artifact, it should be 
called the finite-($T$,$\mu_{\rm Re}$,$\mu_{\rm Im}$) QCD phase diagram.
Anyhow, we have shown that we can apply the KW fermion to in-medium lattice QCD.
We lastly discuss the real-type FCP fermions \cite{misumi}.
As shown in Sec.~\ref{sec:MDF}, we can also consider the 
real-type KW fermion, which loses $\gamma_{5}$ hermiticity. 
We can perform the strong-coupling QCD analysis for this type in a parallel way, 
and can derive the QCD phase diagram as long as the relevant parameter is set to the 
physical range. In the practical lattice QCD simulations, however, we should encounter a
severe sign problem with the real-type FCP fermion even for zero-density cases.
We need further study to judge its applicability to lattice QCD.

\section{Summary and Discussion}
\label{sec:SD}

In this paper we propose a new framework for investigating
the two-flavor finite-($T$,$\mu$) QCD phase diagram. 
We show that the discrete symmetries of the Karsten-Wilczek (KW) fermion 
strongly suggest its applicability to the in-medium lattice QCD. 
To support our idea, we study the strong-coupling lattice QCD
in the medium and derive the phase diagram of chiral symmetry 
for finite temperature and chemical potential.
We have obtained the phase diagram with 1st, 2nd-order and 
crossover critical lines, which is qualitatively in agreement to 
results from the model study.
We also argue that the additive chemical potential renormalization 
to the two flavors can be controlled by adjusting the parameter of 
the dimension-3 counterterm.

In Sec.~\ref{sec:MDF} we review the Karsten-Wilczek fermion and its discrete symmetries.
By the careful comparison between the KW fermion and naive lattice fermions with 
chemical potential, we find that both possess the same discrete symmetries, 
and that KW fermion is a proper formulation for finite-($T$,$\mu$) lattice QCD study. 
It is natural since the KW fermion decouples 14 doublers by assigning them $O(1/a)$ 
imaginary chemical potential, which also leads to the additive chemical potential
renormalization for the rest 2 flavors in the interacting theory. We discuss that, 
in order to control this effective imaginary chemical potential, we need to 
keep a value of the relevant parameter $\mu_{3}$ within 
the minimal-doubling range in Fig.~\ref{AL}. 
It is more modest but similar tuning to the mass parameter tuning in Wilson fermion.
One possible indicator of the parameter ranges is the pion spectrum:
If $\mu_{3}$ is in the no-flavor range, there is no SSB of chiral symmetry
and no massless pion. If $\mu_{3}$ gets into the six-flavor region, 
the number of pseudo Nambu-Goldstone bosons increases.
In Sec.~\ref{sec:SLQ} we perform the strong-coupling finite-($T$,$\mu$) lattice QCD
with the KW fermion. We derive the effective potential of $\sigma$ and $\pi_4$ as a 
function of temperature $T$, chemical potential $\mu$,  and $\mu_{3}$.
As long as keeping $\mu_{3}$ within the physical range in Fig.~\ref{AL}, 
we successfully obtain the QCD chiral phase diagram, which is consistent with
the phenomenological predictions: For high temperature or large chemical potential 
the chiral symmetry restores with chiral condensate's disappearing.
Our result strongly suggests that the Karsten-Wilczek fermion, or more generally 
flavored-chemical-potential fermions are useful for the 2-flavor lattice QCD with 
temperature and density.

One potential problem for this formulation is that, even if we can set $\mu_{3}$ in
the minimal-doubling range, the rest 14 species with infinitely large chemical 
potential may contribute to the thermodynamical quantities.
In such a case we need to subtract a divergent part properly.
Further study is needed to figure out this problem.

Finally we comment on the lattice QCD with this formulation in the small 
chemical potential limit. The minimal-doubling fermion breaks the spatial 
symmetry even for a zero chemical potential limit unless we fine-tune two more 
parameters to restore the hypercubic symmetry \cite{CW}.
We thus expect that this formulation could be less effective to describe
the two-flavor QCD for smaller chemical potential although it works for the 
region near critical lines with sufficiently large chemical potential.


\begin{acknowledgments}
TM is thankful to M.~Creutz and F.~Karsch for the fruitful discussion and hearty encouragement.
TM is supported by Grant-in-Aid for the Japan Society for Promotion of Science (JSPS) 
Postdoctoral Fellows for Research Abroad (No.~24-8).
TK is supported by 
Grant-in-Aid for the Japan Society for Promotion of Science (JSPS) 
Postdoctoral Fellows (No.~23-593).
This work is suppported in part by the Grants-in-Aid for Scientific Research from
JSPS (Nos.
09J01226, 
11J00593, 
23340067, 
24340054, 
and
24540271
),
and by the Grant-in-Aid for the global COE program ``The Next Generation
of Physics, Spun from Universality and Emergence'' from MEXT.
This work is based on fruitful discussion in the YIPQS-HPCI workshop ``New-Type of 
Fermions on the Lattice'', Feb.~9--24, 2012 in Yukawa Institute for Theoretical Physics.
The authors are grateful to the organizers for giving them chances to have interest
in the present topics.
\end{acknowledgments}

\end{document}